\documentclass[aps,showpacs,preprintnumbers,amsmath,amssymb,superscriptaddress,floatfix,nofootinbib]{revtex4}
\usepackage{lipsum}
\usepackage{graphicx}
\usepackage{epsfig}
\usepackage{epstopdf}
\usepackage{hyperref}
\usepackage{amsmath}
\usepackage{amsfonts}
\usepackage{amssymb}
\usepackage{setspace}
\usepackage{amsfonts,color}
\usepackage{caption}
\usepackage{natbib}
\usepackage{subfig}
\usepackage{color}
\usepackage{bbold}
\usepackage{bm}
\newcommand{\roundket}[1]{|#1)}

\def\w2{\tilde w^2}
\def\ws2{1}

\begin{document}

%\title{Large-$N_c$ operator analysis for the SU(3) octet-octet baryon contact interactions in chiral effective field theory }
\title{Large-$N_c$ operator analysis of hyperon-nucleon interactions \\ in SU(3) chiral effective field theory }

\author{Xuyang Liu}
%\email{lxy\_gzu2005@126.com}
\affiliation{School of Mathematics and Physics, Bohai University, Liaoning, 121013, China}

\author{Viroj Limkaisang}
%\email{viroj.li@rmuti.ac.th}
\affiliation{Department of Applied Physics, Faculty of Sciences and Liberal arts, Rajamangala University of Technology Isan, Nakhon Ratchasima, 30000, Thailand}

\author{Daris Samart}
\email{daris.sa@rmuti.ac.th}
\affiliation{Department of Applied Physics, Faculty of Sciences and Liberal arts, Rajamangala University of Technology Isan, Nakhon Ratchasima, 30000, Thailand}
\affiliation{Center Of Excellent in High Energy Physics \& Astrophysics, Suranaree University of Technology, Nakhon Ratchasima, 30000, Thailand}
\affiliation{Advanced Materials and Renewable Energy Research Unit, Rajamangala University of Technology Isan, Nakhon Ratchasima, 30000, Thailand}

\author{Yupeng Yan}
%\email{yupeng@sut.ac.th}
\affiliation{School of Physics, Suranaree University of Technology, Nakhon Ratchasima, 30000, Thailand}
\affiliation{Center Of Excellent in High Energy Physics \& Astrophysics, Suranaree University of Technology, Nakhon Ratchasima, 30000, Thailand}

\date{\today}

\vskip 1pc
\begin{abstract}
We study octet-octet baryon ($J^P = {\textstyle \frac12}^+$) contact interactions in SU(3) chiral effective
field theory by using large-$N_c$ operator analysis. Applying the $1/N_c$ expansion of the Hartee Hamiltonian, we find 15 operators in the octet-octet baryon potential where 4 operators for leading order (LO) and 11 for and net-to-next-to-leading order (NNLO). The large-$N_c$ operator analysis of octet-octet baryon matrix elements reduces the number of free parameters from 15 to 6 at LO of the $1/N_c$ expansion. The application of large-$N_c$ sum rules to the J\"{u}lich model of hyperon-nucleon (YN) interactions at the LO of the chiral expansion reduces the model parameters to 3 from 5 at the LO of $1/N_c$ expansion. We find that the values of LECs fitted to YN scattering data in Ref. \cite{Li:2016paq} in the relativistic covariant ChEFT (EG) approach is more consistent with the predictions of large-$N_c$ than in the heavy baryon (HB) formalism approach.
\end{abstract}

\maketitle
\section{Introduction}
%1. Talk about ChEFT and what ChEFT has done in baryon-baryon
%Effective field theory (EFT) plays an important role on non-perturbative QCD. This leads to
Chiral effective field theory (ChEFT) \cite{Weinberg:1978kz,Gasser:1983yg}, based on the approximately and spontaneously broken chiral symmetry of QCD, allows for a systematic way of calculating low-energy hadronic observables. It is very efficient and convenient to use hadrons as basic degrees of freedom rather than quarks and gluons in the ChEFT. Chiral Lagrangian is required to include all possible interactions between hadrons which are constructed in terms of the relevant symmetries of QCD \cite{Scherer:2012xha}. A number of low-energy properties in the strong interaction is very successfully described by using the ChEFT. The ChEFT is also utilized to shed light on the study of nuclear forces (see \cite{Epelbaum:2008ga,Machleidt:2011zz} for reviews). It was demonstrated by Weinberg's seminal works \cite{Weinberg:1990rz,Weinberg:1991um} that one can calculate the nuclear forces systematically by using appropriate power counting scheme. Therefore, loop-corrections and higher order terms can be included for the accuracy of the calculations. Nucleon-nucleon (NN) forces derived in the ChEFT successfully described a huge number of NN experimental data. The NN potentials are composed of the long and short range interactions, where the long range NN force is mainly contributed by the pion exchange while the short range part is encoded by contact term NN interactions with unknown low-energy constants (LECs) to be fitted to experimental data. The higher order contact terms of the NN potentials have been constructed in Refs. \cite{Ordonez:1993tn,Ordonez:1996} at next-to leading order (NLO) and in Refs. \cite{Epelbaum:2004fk,Entem:2003ft} for next-to-next-to-next-to leading order (N$^3$LO) in terms of chiral expansions.

%2. Talk about strangeness and hyper-nuclei
On the other hand, hyperon-nucleon (YN) and hyperon-hyperon (YY) forces have been less studied compared with the NN forces. YN interactions are keys for understanding hyper-nuclei and neutron stars \cite{Nogga:2001ef,Lonardoni:2014bwa}. The contact and meson exchange terms of the YN interactions in the ChEFT were constructed by using the SU(3) flavor symmetry in Ref. \cite{Polinder:2006zh} at leading order (LO) and extended to NLO in Ref. \cite{Haidenbauer:2013oca}. The most general SU(3) chiral Largrangians of the octet-octet baryon contact term interactions have been worked out in Ref. \cite{Petschauer:2013uua}. The study of the YY interactions was performed in Refs. \cite{Polinder:2007mp,Haidenbauer:2015zqb,Haidenbauer:2009qn}. At the LO of the YN interactions \cite{Polinder:2006zh,Li:2016paq}, the SU(3) chiral Lagrangian has 15 free parameters (LECs) and the partial-wave expansion analysis leads to 5 LECs which are fixed with YN data. In this work, we will use the large-$N_c$ operator analysis to explore the $N_c$ scales and reduce the number of the unknow LECs in the SU(3) chiral Largrangians and in the LO YN potential \cite{Polinder:2006zh,Li:2016paq}. %The SU(3) baryon-baryon potential has a lot more LECs, and less data, so using large-$N_c$ constraints to help the fitting may be very useful.

%3. Large-Nc in ChEFT and baryon-baryon
Large-$N_c$ is an approximate framework of QCD and very useful in the study of hadrons at low-energies. The basic idea is that one can consider the number of colors ($N_c$) to be large and expand it in power of $1/N_c$ \cite{'tHooft:1973jz,Witten:1979kh}. By using this framework, a number of simplifications of QCD occurs in the large-$N_c$ limit (see Refs. \cite{Jenkins:1998wy,Matagne:2014lla} for reviews). The $1/N_c$ expansion of QCD for the baryon \cite{Dashen:1993jt,Dashen:1994qi,Luty:1993fu} has been applied to the NN potential in \cite{Kaplan:1995yg,Kaplan:1996rk,Banerjee:2001js} and three-nucleon potential in \cite{Phillips:2013rsa}. Moreover, the $1/N_c$ expansion is used to study parity-violating NN potentials in \cite{Phillips:2014kna,Schindler:2015nga} as well as time-reversal violating NN potentials \cite{Samart:2016ufg}. The study of the large-$N_c$ analysis in the NN system provides the understanding of the $N_c$ scales of the LECs in the NN forces. In addition, the $1/N_c$ expansion also helps us to reduce the independent number of the LECs \cite{Schindler:2015nga}. However, the octet-octet baryon interactions in SU(3) flavor symmetry have not been investigated in the large-$N_c$ approach. In this work, we will extend the large-$N_c$ operator analysis in Refs. \cite{Kaplan:1996rk,Phillips:2013rsa} to the SU(3) chiral Lagrangian in Refs. \cite{Polinder:2006zh,Li:2016paq}. The large-$N_c$ octet-octet baryon potential is constructed up to NNLO in terms of the $1/N_c$ expansion. %Comparing to the octet-octet baryon SU(3) chiral potential, we find 6 free parameters (the independent number of the LECs) at LO of $1/N_c$.
We will apply large-$N_c$ sum rules to YN interactions at LO which has been recently investigated in Ref. \cite{Li:2016paq}.
%The large-$N_c$ sum rules of this work can be tested with the LECs of YN interactions at the LO where the results of the YN interactions
%from two approaches that have been done in \cite{Li:2016paq}.
%The results of this work are also useful for fitting the LECs to the YN scattering data with less number of the free parameters that constraint from the large-$N_c$ arguments.
Moreover, the results can be applied to the YN at NLO and YY sector.

%4. Organize the the paper
We outline this work as follows: In section 2 we will setup the matrix elements of the octet-octet baryon potential from the SU(3) chiral Lagrangian. In the next section, the potential of the $1/N_c$ expansion is constructed up to NNLO and large-$N_c$ sum rules for LECs are implied. In section 4, we apply results of the large-$N_c$ sum rules to the LO YN potential. In the last section, we give the conclusion in this work.

\section{The potential of the SU(3) octet-octet baryon contact term interactions}
We start with the SU(3) chiral Largrangian of the octet-octet baryon interactions and it was proposed by Ref. \cite{Polinder:2006zh}.
The SU(3)-flavor symmetry is imposed and the chiral Lagrangian is Hermitian and invariant under Lorentz transformations and the CPT discrete symmetry is implied. The minimal SU(3) invariant chiral Lagrangian with non-derivative is given by,
\begin{eqnarray}\label{chi-L}
{\mathcal L}^{(1)} &=& C^{(1)}_i \left<\bar{B}_1\bar{B}_2\left(\Gamma_i B\right)_2\left(\Gamma_i B\right)_1\right>\ , \nonumber \\
{\mathcal L}^{(2)} &=& C^{(2)}_i \left<\bar{B}_1\left(\Gamma_i B\right)_1\bar{B}_2\left(\Gamma_i B\right)_2\right>\ , \nonumber \\
{\mathcal L}^{(3)} &=& C^{(3)}_i \left<\bar{B}_1\left(\Gamma_i B\right)_1\right>\left<\bar{B}_2\left(\Gamma_i B\right)_2\right>\  .
\end{eqnarray}
Here $1$ and $2$ denote the label of the particles in the scattering process, the $B$ is the usual irreducible octet representation of SU(3) given by
\begin{eqnarray}
B&=& \frac{1}{\sqrt 2}\sum_{a=1}^8 \lambda^a B^a =
\left(
\begin{array}{ccc}
\frac{\Sigma^0}{\sqrt{2}}+\frac{\Lambda}{\sqrt{6}} & \Sigma^+ & p \\
\Sigma^- & \frac{-\Sigma^0}{\sqrt{2}}+\frac{\Lambda}{\sqrt{6}} & n \\
-\Xi^- & \Xi^0 & -\frac{2\Lambda}{\sqrt{6}}
\end{array}
\right) \ ,
\label{eq:7}
\end{eqnarray}
where the $\langle \cdots \rangle$ brackets denote taking the trace in the three-dimensional flavor space and the normalization of Gell-Mann matrices $\langle \lambda^a\,\lambda^b \rangle = 2\,\delta^{ab}$ is used. The $\Gamma_i$ are the usual elements of the Clifford algebra
\begin{equation}
\Gamma_1=1 \, , \,\,
\Gamma_2=\gamma^\mu \, , \,\,
\Gamma_3=\sigma^{\mu\nu} \, , \,\,
\Gamma_4=\gamma^\mu\gamma_5  \, , \,\,
\Gamma_5= i\,\gamma_5 \,\, .
\label{eq:2.2}
\end{equation}
By using the chiral power counting in Ref. \cite{Polinder:2006zh}, it has been shown that we have 15 LO non-derivative terms of the chiral Lagrangian.
It has also been demonstrated in Ref. \cite{Polinder:2006zh} that the above Lagrangians are the minimal set of the contact interaction terms in terms of flavor and spin structures by using Cayley-Hamilton identity and Fierz transformation.

To obtain the potentials, we follow approach in Refs. \cite{Girlanda:2010ya,Girlanda:2010zz} by imposing relativistic covariant constraints. Letting  ${\mathcal H} = -\,{\mathcal L}$ and taking the approach of the relativistic constraints in \cite{Girlanda:2010ya,Girlanda:2010zz} into account, one obtains the potential of the octet-octet baryon contact interactions up to the second order of the small momenta of the baryons and it reads,
\allowdisplaybreaks
\begin{eqnarray}\label{pot-1}
V^{(1)}&=& \langle\bar\chi_2, d\,;\, \bar\chi_1, c\,|\, {\mathcal H}^{(1)} |\, a,\chi_1\,;\, b,\chi_2 \rangle
\nonumber\\
&=& \left\{ \frac13\,\delta^{cd}\delta^{ba} + \frac12\,\big(\,d^{cde} +if^{cde} \big)\big( d^{eba} + if^{eba}\big) \right\}
\nonumber\\
&&\quad \times\,\Big\{\, c_S^{(1)}\tilde O_S + c_T^{(1)}\tilde O_T
+ \left(c_1^{(1)}p_-^2 + c_2^{(1)}p_+^2 \right)\delta_{\bar\chi_1\chi_1}\delta_{\bar\chi_2\chi_2} + \left( c_3^{(1)}p_-^2 + c_4^{(1)}p_+^2 \right)\vec\sigma_1 \cdot \vec\sigma_2
\nonumber\\
&&\qquad\qquad\qquad  +\, c_5^{(1)}\frac{i}{2} (\vec\sigma_1 + \vec\sigma_2) \cdot \left(\vec p_+\times\vec p_- \right) + c_6^{(1)}(\vec p_-\cdot\vec\sigma_1)(\vec p_-\cdot\vec\sigma_2) + c_7^{(1)}(\vec p_+\cdot\vec\sigma_1)(\vec p_+\cdot\vec\sigma_2)  \,\Big\},
\end{eqnarray}
where
\begin{eqnarray}
\tilde O_S &=& \delta_{\bar\chi_1\chi_1}\delta_{\bar\chi_2\chi_2} + \frac{i}{2M^2} \left(\vec p_+\times\vec p_- \right) \cdot (\vec\sigma_1 - \vec\sigma_2)\,,
\nonumber\\
\tilde O_T &=& \vec\sigma_1 \cdot \vec\sigma_2 - \frac{i}{2M^2} \left(\vec p_+\times\vec p_- \right) \cdot (\vec\sigma_1 - \vec\sigma_2) \,,
\end{eqnarray}
and $\vec \sigma_{i}\equiv \vec\sigma_{\bar\chi_i\chi_i}$ with $i=1,2$. The indices $a\,(c)$, $b\,(d)$, $\chi_1\,(\bar\chi_{1})$ and $\chi_2\,(\bar\chi_{2})$ are flavor and spin indices of incoming (outgoing) baryon number 1 and 2 respectively and $M$ is the octet baryon mass in the SU(3) flavor symmetry limit. We note that the octet-octet baryon potentials agree with the heavy baryons formulation of ChEFT in \cite{Pastore:2009is,Epelbaum:1998ka} for the spin structures.
By using the partial integrations and the baryon equation of motion to eliminate time derivative as shown in Refs. \cite{Girlanda:2010ya,Girlanda:2010zz}, the potential in Eq. (\ref{pot-1}) is the minimal set of linearly independent operators and it consists of 2 LO and 7 NLO operators (see appendix \ref{appA} for the detail derivation of the potential). The LECs, $c_i^{(1)}$ are linear combinations of the couplings $C_i^{(1)}$ as,
\begin{eqnarray}\label{new-LECs-1}
c_S^{(1)} &=& C_1^{(1)} + C_2^{(1)} \,,
\qquad
c_T^{(1)} = C_3^{(1)} - C_4^{(1)} \,,
\qquad
c_1^{(1)} = -\frac{1}{4M^2}\left( C_2^{(1)} + C_3^{(1)} \right),\qquad
c_2^{(1)} = -\frac{1}{2M^2}\left( C_1^{(1)} - C_2^{(1)} \right),
\nonumber\\
c_3^{(1)} &=& -\frac{1}{4M^2}\left( C_2^{(1)} + C_3^{(1)} \right),\qquad
c_4^{(1)} = \frac{1}{4M^2}\left( C_3^{(1)} - C_4^{(1)} \right), \qquad
c_5^{(1)} = -\frac{1}{2M^2}\left( C_1^{(1)} - 3 C_2^{(1)} - 3 C_3^{(1)} - C_4^{(1)}  \right),
\nonumber\\
c_6^{(1)} &=& \frac{1}{4M^2}\left( C_2^{(1)} + C_3^{(1)} +  C_4^{(1)} + C_5^{(1)} \right), \qquad
c_7^{(1)} = -\frac{1}{4M^2}\left( C_3^{(1)} + C_4^{(1)} \right).
\end{eqnarray}
In addition, it is worth to discuss about the chiral power counting ($Q/M$) where a $Q$ is typical three momentum of the baryon. %If we specify the counting of the baryon mass, M as follow,
If we impose $M \sim \Lambda$ where $\Lambda$ is a chiral symmetry breaking scale. Therefore, our power counting rule in this work adopts $Q/M \sim \left(Q/\Lambda\right)^2$ which has been used in Refs. \cite{Epelbaum:2004fk,Ordonez:1996} for the NN potentials. The notations of the momentum in this work are defined below
\begin{eqnarray}\label{momentum-pm}
\vec p_+ = \frac12(\vec p\,' + \vec p)\,,\quad p_+^2 = \vec p_+\cdot\vec p_+\,,\qquad \vec p_- = \vec p\,' - \vec p\,,\quad p_-^2 = \vec p_-\cdot\vec p_-\,,
\qquad \vec n = \vec p\times\vec p\,' = \vec p_+\times\vec p_-\,,
\end{eqnarray}
where $\vec p\,(\vec p\,'\,)$ is incoming (outgoing) three-momentum in the c.m. frame and the on-shell condition of the external momenta is given by
\begin{eqnarray}
\vec p_+\cdot\vec p_- = 0\,.
\end{eqnarray}
With the same manner, the octet-octet baryon potentials for $C_i^{(2)}$ and $C_i^{(3)}$ are written by
\allowdisplaybreaks
\begin{eqnarray}\label{pot-2}
V^{(2)} &=& \langle\bar\chi_2, d\,;\, \bar\chi_1, c \,|\, {\mathcal H}^{(2)} \,|\; a,\chi_1\,;\, b,\chi_2 \rangle
\nonumber\\
&=& \left\{ \frac13\,\delta^{ca}\delta^{bd} + \frac12\,\big(\,d^{cae} +if^{cae} \big)\big( d^{edb} + if^{edb}\big) \right\}
\nonumber\\
&&\quad \times\,\Big\{\, c_S^{(2)}\tilde O_S + c_T^{(2)}\tilde O_T
+ \left(c_1^{(2)}p_-^2 + c_2^{(2)}p_+^2 \right)\delta_{\bar\chi_1\chi_1}\delta_{\bar\chi_2\chi_2} + \left( c_3^{(2)}p_-^2 + c_4^{(2)}p_+^2 \right)\vec\sigma_1 \cdot \vec\sigma_2
\nonumber\\
&&\qquad\qquad\qquad  +\, c_5^{(2)}\frac{i}{2} (\vec\sigma_1 + \vec\sigma_2) \cdot \left(\vec p_+\times\vec p_- \right) + c_6^{(2)}(\vec p_-\cdot\vec\sigma_1)(\vec p_-\cdot\vec\sigma_2) + c_7^{(2)}(\vec p_+\cdot\vec\sigma_1)(\vec p_+\cdot\vec\sigma_2)  \,\Big\}\,,
\end{eqnarray}
and
\allowdisplaybreaks
\begin{eqnarray}\label{pot-3}
&& V^{(3)} = \langle\bar\chi_2, d\,;\, \bar\chi_1, c \,|\, {\mathcal H}^{(3)} \,|\; a,\chi_1\,;\, b,\chi_2 \rangle
\nonumber\\
&&\qquad\, = \delta^{ca}\delta^{bd}\Big\{\, c_S^{(3)}\tilde O_S + c_T^{(3)}\tilde O_T
+ \left(c_1^{(3)}p_-^2 + c_2^{(3)}p_+^2 \right)\delta_{\bar\chi_1\chi_1}\delta_{\bar\chi_2\chi_2} + \left( c_3^{(3)}p_-^2 + c_4^{(3)}p_+^2 \right)\vec\sigma_1 \cdot \vec\sigma_2
\nonumber\\
&&\qquad\qquad\qquad\qquad\qquad  +\, c_5^{(3)}\frac{i}{2} (\vec\sigma_1 + \vec\sigma_2) \cdot \left(\vec p_+\times\vec p_- \right) + c_6^{(3)}(\vec p_-\cdot\vec\sigma_1)(\vec p_-\cdot\vec\sigma_2) + c_7^{(3)}(\vec p_+\cdot\vec\sigma_1)(\vec p_+\cdot\vec\sigma_2)  \,\Big\}\,,
\end{eqnarray}
where the LECs in Eqs. (\ref{pot-2}) and (\ref{pot-3}) are the linear combinations of the couplings as in Eq. (\ref{new-LECs-1}) by replacing $c_i^{(1)}\rightarrow c_i^{(2,3)}$ and $C_i^{(1)}\rightarrow C_i^{(2,3)}$\,. By using relativistic reductions as in \cite{Girlanda:2010ya,Girlanda:2010zz}, we obtain the minimal set of the SU(3) octet-octet baryon potentials and there are 27 operators totally. Moreover, Fierz identities for the Gell-mann matrices ($\lambda^a$) are also taken into account for the calculations of the potentials in Eqs. (\ref{pot-1}), (\ref{pot-2}) and (\ref{pot-3}). We found that there is no the redundant terms of the SU(3) flavor structures. We obtain 6 and 21 operators at LO and NLO of the small momentum scale expansion ($Q/M$). At the LO, the operators from the couplings $C_{1,2,3,4}^{(1,2,3)}$ enter to contribute the potential but the couplings $C_5^{(1,2,3)}$ start at NLO. %One notes that since we adopted $M\sim \Lambda$ for the counting rule. The potentials of the SU(3) chiral Lagrangian baryon contact terms interaction take the forms as NLO which are similar forms in Ref. \cite{Haidenbauer:2013oca}. But the potentials in this work have additional term in the leading order interactions. This is because of the relativistic covariant constraints as well as the partial integrations of the baryon's equation of motion and the Friez identities are taken into account as shown in \cite{Girlanda:2010ya,Girlanda:2010zz}.
%\begin{eqnarray}
%\lambda_{\alpha\beta}^a\,\lambda_{\gamma\delta}^b &=& \frac13\,\lambda_{\alpha\delta}^a\,\lambda_{\gamma\beta}^b + \frac12\,\big( \lambda^a\lambda^c\big)_{\alpha\delta}\big(\lambda^b\lambda^c\big)_{\gamma\beta}
%\nonumber\\
%\big( \lambda^a\lambda^c\big)_{\alpha\beta}\big(\lambda^b\lambda^c\big)_{\gamma\delta} &=& \frac{16}{9}\,\lambda_{\alpha\delta}^a\,\lambda_{\gamma\beta}^b -\frac13\,\big( \lambda^a\lambda^c\big)_{\alpha\delta}\big(\lambda^b\lambda^c\big)_{\gamma\beta}\,,
%\end{eqnarray}
%where Greek alphabets $\alpha,\beta,\gamma\,\delta$ are fundamental representation indices.
We will reduce the independent number of the LECs of the SU(3) octet-octet baryon interactions in the ChEFT by using the large-$N_c$ operator analysis in the next section.

\section{The $1/N_c$ operator product expansion analysis of the two-baryon matrix elements}
\subsection{The $1/N_c$ expansion octet-octet baryon ansatz}
In this section, we are going to study the $1/N_c$ expansion for the octet-octet baryon matrix elements. According to Witten's conjecture \cite{Witten:1979kh}, the matrix elements of baryon-baryon scattering should scale like $N_c$, i.e. \cite{Kaplan:1995yg,Kaplan:1996rk}\,,
\begin{eqnarray}
N_c\big\langle B_1\,|\,\mathcal{\hat O}_1^{i}\,|\, B_{1} \big\rangle \big\langle B_2\,|\,\mathcal{\hat O}_2^{i'}\,|\, B_{2} \big\rangle\,,
\end{eqnarray}
where $\mathcal{\hat O}_1^{i}$ and $\mathcal{\hat O}_2^{i'}$ operators are the $i$- and $i'$-quark current operators on the first and the second baryon. It has proven in the Ref. \cite{Luty:1993fu} that the matrix elements for one baryon in SU(3) flavor symmetry has the $N_c$ scaling as,
\begin{eqnarray}
\big\langle B_j\,|\,\mathcal{\hat O}_j^{i}\,|\, B_j \big\rangle \lesssim N_c^0\,,
\end{eqnarray}
with $j=1,~2$\,. This holds for the matrix elements of the second baryon as well. One can expand the matrix elements in terms of effective quark operator and effective spin-flavor baryon states in $1/N_c$ expansion as \cite{Dashen:1994qi,Luty:1993fu},
\begin{eqnarray}
\big\langle B\,|\,\mathcal{\hat O}^{i}\,|\, B \big\rangle = \big( B\,|\,\sum_r c_r^{(i)}\left(\frac{\mathcal{O}}{N_c}\right)^r\,|\, B \big)\,,
\end{eqnarray}
where $c_r^{(i)}$ is a function which contains dynamical properties of the system and $|\, B \big)$ is an effective baryon state composed of spin and flavor structures only \cite{Dashen:1994qi,Luty:1993fu}. The $\mathcal{O}^r$ are the $r$-body operators which comprises of the effective quark operators \cite{Kaplan:1995yg,Kaplan:1996rk},
\begin{eqnarray}
\left(\frac{\mathcal{O}}{N_c}\right)^r = \left(\frac{J}{N_c}\right)^l\,\left(\frac{T}{N_c}\right)^m\,\left(\frac{G}{N_c}\right)^n\,,\quad {\rm with}\quad \,l+m+n = r\,.
\end{eqnarray}
The operators $J$\,, $T$\, and $G$\, are spin, flavor and spin-flavor operators, respectively and they are defined by \cite{Dashen:1994qi,Lutz:2010se},
\begin{eqnarray}
&& \mathbb{1} = q^\dagger ( \mathbf{1}  \otimes \mathbf{1} )\,q \,, \qquad  \qquad \;\;
J_i = q^\dagger \Big(\frac{\sigma_i }{2} \otimes \mathbf{1})\, q \,,
\nonumber\\
&& T^a = q^\dagger \Big(\mathbf{1} \otimes \frac{\lambda_a}{2} \Big)\, q\, ,\qquad \quad \;\;
G^a_i = q^\dagger \Big( \frac{\sigma_i}{2} \otimes \frac{\lambda_a}{2} \Big)\, q\,,
\label{def:one-body-operators}
\end{eqnarray}
where $q$ and $q^\dagger$ are quark annihilation and creation operators respectively. According to the fully antisymmetry and Fermi statistics of the SU($N_c$) color group, the spin and flavor of baryonic ground state of the $N_c$ quarks have to be completely symmetric representation. Therefore one can consider quark operators $q$ and $q^\dagger$ as bosonic operators with the commutation relation $\left [ q\,,\, q^\dagger\right] = 1$\,. The $N_c$ scaling of the $r$-body operator $\mathcal{O}^r$ and the the coefficient $c_r^{(i)}$ scale like \cite{Kaplan:1995yg,Kaplan:1996rk},
\begin{eqnarray}
\big( B\,|\,\mathcal{O}^r \,|\, B \big) \lesssim  N_c^r\,,\qquad c_r^{(i)} \sim N_c^0\,.
\end{eqnarray}
In addition, The one-baryon matrix elements of the operators $J$\,, $T$\, and $G$ in SU(3) flavor symmetry framework have $N_c$ scaling in the following way \cite{Dashen:1994qi}
\begin{eqnarray}\label{Nc-scale-operators}
\big(B\,|\,J^i\,|\,B\big) \sim N_c^0 \,,\quad \big(B\,|\,\mathbb{1}\,|\,B\big) \sim N_c\,,\quad \big(B\,|\,T^a\,|\,B\big) \lesssim N_c\,,\quad \big(B\,|\,G^{i\,a}\,|\,B\big) \lesssim N_c\,.
\end{eqnarray}
In contrast to the SU(2) flavor symmetry, there is only one operator that can suppress rising of the $N_c$ for one-baryon matrix elements i.e. the $J$ whereas all the rest of the effective operators rises the $N_c$ factor. However, the symbol, $\lesssim$ is used for saturating the maximum of the $N_c$ scaling for the $\big(B\,|\,T^a\,|\,B\big)$ and $\big(B\,|\,G^{i\,a}\,|\,B\big)$ because the matrix elements of the $T^a$ operator scales like $N_c^0$ for $a=1,2,3$, but as $\sqrt{N_c}$ when $a=4,5,6,7$ and as $N_c$ when $a=8$\,. On the other hand, the matrix elements of the $G^{i\,a}$  scales like $N_c$ for $a=1,2,3$, as $\sqrt{N_c}$ when $a=4,5,6,7$ and as $N_c^0$ when $a=8$ \cite{Dashen:1994qi}. These are the differences of the effective operators between SU(2) and SU(3) flavor symmetries. Moreover, it is worth to discuss about the $N_c$ scaling of the external momentum variables. Here we consider all momentum in c.m. frame as we discussed in the previous section. One recalls the $N_c$ scaling of the momentum variables in Eq. (\ref{momentum-pm}), it reads \cite{Kaplan:1996rk},
\begin{eqnarray}\label{momentum-Nc}
\vec p_+ \sim 1/N_c\,,\qquad \vec p_- \sim N_c^0\,.
\end{eqnarray}
In a meson exchange picture, the $\vec p_+$ can only appear in the baryon-baryon potential as a relativistic correction (i.e., a velocity dependent term). Therefore, the $\vec p_+$ always come with the factor $1/M$\,. Since $M\,\sim\,N_c$, this gives $\vec p_+ \sim 1/N_c$ (for more detail discussions see \cite{Kaplan:1996rk,Phillips:2013rsa,Schindler:2015nga}).
%The momentum $\vec p_+$ has $N_c$ scaling dependence in Eq. (\ref{momentum-Nc}) because  as a relativistic correction in $t$-channel meson exchange picture i.e. it is velocity dependent term,
The baryon-baryon potential in terms of $1/N_c$ expansion can be written in the Hartee Hamiltonian \cite{Kaplan:1996rk,Phillips:2013rsa,Schindler:2015nga}. It takes the following form,
\begin{eqnarray}\label{hartee}
\hat H = N_c\sum_{r}\sum_{lm} c_{r,lm}\, \left(\frac{J}{N_c}\right)^l \left(\frac{T}{N_c}\right)^m \left(\frac{G}{N_c}\right)^{r-l-m}\,,
\end{eqnarray}
where again the $c_{r,lm}$ coefficient function has scale $N_c^0$\,. It is well know that, at the large-$N_c$ limit, the spin-$1/2$ and $3/2$ baryons are degeneracy states. In this work, we project the Hamiltonian $\hat H$ to the octet (spin-$1/2$) baryon sector only. This has been discussed extensively in \cite{Kaplan:1996rk}.
%The relevant $1/N_c$ potential ansatz of the baryon-baryon scattering in contact term is ready to construct.
We will construct the Hamiltonian in order of $1/N_c$ expansion.
Then the leading-order (LO) is given by
\begin{eqnarray}\label{LO}
\hat H_{\rm LO} &=& U_1^{\rm LO}(p_-^2)\, \mathbb{1}_1\cdot\mathbb{1}_2 + U_2^{\rm LO}(p_-^2)\, T_1\cdot T_2 + U_3^{\rm LO}(p_-^2)\, G_1\cdot G_2
+ U_4^{\rm LO}(p_-^2)\, (p_-^i p_-^j)_{(2)}\cdot (G_{1}^{i,a} G_{2}^{j,a})_{(2)} \,,
\end{eqnarray}
where $T_1\cdot T_2 = T_1^a T_2^a$ and $G_1\cdot G_2 = G_1^{i,a} G_2^{i,a}$\,. $U_i^{\rm LO}(p_-^2)$ is arbitrary function of the $p_-^2$ and it has $N_c^0$ scale. Here we also introduce the notation,
\begin{eqnarray}
(A^i B^j)_{(2)} \equiv \frac12\left( A^iB^j + A^jB^i - \frac23\delta_{ij}A\cdot B\right),
\end{eqnarray}
and then
\begin{eqnarray}
(p_\pm^i p_\pm^j)_{(2)}\cdot(\sigma_1^i\sigma_2^j)_{(2)} = (\vec p_\pm\cdot\vec\sigma_1)( \vec p_\pm\cdot\vec\sigma_2) - \frac13\,p_\pm^2\sigma_1\cdot\sigma_2 \,.
\end{eqnarray}
%Next we construct the octet-octet baryon Hamiltonian at Next-to-Leading-Order (NLO), it reads,
%\begin{eqnarray}\label{NLO}
%\hat H_{\rm NLO} = U_1^{\rm NLO}(p_-^2)\,i\, (\vec p_+\times\vec p_-)\cdot(T_1^a \vec G_2^a + \vec G_1^a T_2^a)\,,
%\end{eqnarray}
%where $\vec G^a \equiv G^{i,a}$\,.
In this work, we terminate the $1/N_c$ expansion at the $1/N_c^2$ order. Then, the octet-octet baryon Hamiltonian at NNLO takes the following form,
\begin{eqnarray}\label{NNLO}
\hat H_{\rm NNLO} &=&  U_1^{\rm NNLO}(p_-^2)\, p_+^2 \mathbb{1}_1\cdot\mathbb{1}_2 + U_2^{\rm NNLO}(p_-^2)\, \vec J_1\cdot\vec J_2 + U_3^{\rm NNLO}(p_-^2)\,\vec J_1\cdot\vec J_2\,T_1\cdot T_2 + U_4^{\rm NNLO}(p_-^2)\, p_+^2 T_1\cdot T_2
\nonumber\\
&+& U_5^{\rm NNLO}(p_-^2)\, p_+^2 G_1\cdot G_2 + U_6^{\rm NNLO}(p_-^2)\, i\,(\vec p_+ \times \vec p_-)\cdot(\vec J_1 + \vec J_2)
+ U_7^{\rm NNLO}(p_-^2)\,i\, (\vec p_+\times\vec p_-)\cdot(T_1^a \vec G_2^a + \vec G_1^a T_2^a)
\nonumber\\
&+& U_8^{\rm NNLO}(p_-^2)\, i\,(\vec p_+ \times \vec p_-)\cdot(\vec J_1 + \vec J_2)\, T_1\cdot T_2
+ U_9^{\rm NNLO}(p_-^2)\, (p_-^ip_-^j)_{(2)}\cdot(J_1^{i}J_2^{j})_{(2)}
\nonumber\\
&+& U_{10}^{\rm NNLO}(p_-^2)\,(p_-^ip_-^j)_{(2)}\cdot(J_1^{i}J_2^{j})_{(2)}\, T_1\cdot T_2 + U_{11}^{\rm NNLO}(p_-^2)\, (p_+^ip_+^j)_{(2)}\cdot(G_1^{i,a}G_2^{j,a})_{(2)}\,.
\end{eqnarray}
Here the $1/N_c$ scale factor is implied on each effective operators, $\mathbb{1}$, $J$, $T$ and $G$ implicitly. The functions $U_{i}^{\rm LO}(p_-^2)$ and $U_{i}^{\rm NNLO}(p_-^2)$ have $N_c^0$ scale. Noting that there are no $p_+^2 J_1\cdot J_2$ and $(p_+^ip_+^j)_{(2)}\cdot(J_1^{i}J_2^{j})_{(2)}$ structures because these operators have a further suppression in order $1/N_c^4$\,.

Let's us discuss comparisons between the octet-octet baryon potential and the nucleon-nucleon potential in the $1/N_c$ expansion. In the case of the SU(3) flavor symmetry, we find addition operator $T_1\cdot T_2$ at LO instead of NNLO because $T^8\,T^8/N_c \,\sim\,N_c$\, while there is no such operator in nucleon-nucleon potential. Superficially, the two-body operator, $T^a G^{i\,a}/N_c$ should scale like $N_c$ by using the $N_c$ scale counting rules in Eq. (\ref{Nc-scale-operators}). But if we consider the operator more carefully then we find $T^a G^{i\,a}/N_c \sim N_c^0$ because $T^{1,2,3} G^{i\,1,2,3}/N_c \,\sim$ $T^{4,5,6,7} G^{i\,4,5,6,7}/N_c \,\sim$ $T^{8} G^{i\,8}/N_c \,\sim\,N_c^0$\,. Surprisingly, the SU(3) octet-octet potential has the same structures as the nucleon-nucleon potential in SU(2) flavor symmetry i.e. there is no NLO term in the $1/N_c$ expansion. The extension of the flavor symmetry from SU(2) to SU(3) does not change the profile of the $1/N_c$ potential.  Before closing this section, we would like to summarize the $1/N_c$ expansion octet-octet baryon Hamiltonian. There are 4 LO operators. At the NNLO of $1/N_c$ expansion, we obtain 11 operators. We totally have 15 operators of $1/N_c$ expansion for octet-octet baryon potential.

\subsection{Matching the octet-octet baryon potential of the SU(3) chiral Lagrangian with the $1/N_c$ operator product expansion}
We will evaluate, in this section, the octet-octet baryon potential from the Hartee Hamiltonian in Eqs. (\ref{LO}) and (\ref{NNLO}). The $1/N_c$ potential is given by
\begin{eqnarray}
V = \big(\bar\chi_{2},d\,;\,\bar\chi_{1},c\,|\,\hat H\,|\,a,\chi_1\,;\,b,\chi_2\big),
\end{eqnarray}
where $a\,(c)$, $b\,(d)$, $\chi_1\,(\bar\chi_{1})$ and $\chi_2\,(\bar\chi_{2})$ are flavor and spin indices of incoming (outgoing) baryon number 1 and 2 respectively.
After that we will do matching the octet-octet baryon potential and $1/N_c$ operator product expansion to correlate the LECs from the chiral Lagrangian in Eq. (\ref{chi-L}). First of all, we recall the action of the effective operators on the effective baryon states at $N_c=3$ as \cite{Lutz:2010se},
\begin{eqnarray}\label{one-body}
\mathbb{1} \,\roundket{a,\chi}&=& 3\, \roundket{a,{\bar \chi}}\,,
\nonumber \\
J_i \,\roundket{a,\chi}&=&\frac{1}{2}\, \sigma^{(i)}_{{\bar \chi} \chi}\, \roundket{a,{\bar \chi}}\,,
\nonumber \\
T^a\, \roundket{b,\chi}&=& i\,f^{bca}\, \roundket{c,\chi}\,,
\nonumber \\
G^{a}_i\, \roundket{b,\chi}&=&  \sigma^{(i)}_{{\bar \chi} \chi}\, \Big(\frac12\,d^{bca} + \frac{i}{3}\, f^{bca}\Big)\,
\roundket{c,{\bar \chi}} + \,\cdots\,,
\end{eqnarray}
where $\cdots$ stands for a relevant structure of spin-${\textstyle \frac32}$ baryons \cite{Lutz:2010se} but we do not consider the spin-${\textstyle \frac32}$ baryons degree of freedom in this work. Before matching operators, we make ansazt for the arbitrary functions $U_i^{LO}$ and $U_i^{NNLO}$ that they are,
\begin{eqnarray}
U_i^{LO}(p_-^2) = g_i\,,\qquad U_i^{NNLO}(p_-^2) = h_i\,.
\end{eqnarray}
Using Eq. (\ref{one-body}) in Eqs. (\ref{LO}) and (\ref{NNLO}), the potential in terms of the large-$N_c$ operators at the LO is given by,
\begin{eqnarray}\label{LO-pot}
V_{\rm LO} &=& 9\,g_1\,\delta_{\bar\chi_1\chi_1}\delta_{\bar\chi_2\chi_2}\,\delta^{cd}\delta^{bd}
+ g_2\,i^2\,f^{ace}\,f^{bde}\,\delta_{\bar\chi_1\chi_1}\delta_{\bar\chi_2\chi_2}
+ g_3\,\vec\sigma_1\cdot\vec\sigma_2\,\big( {\textstyle \frac12}\,d^{ace} + {\textstyle \frac{i}{3}}\,f^{ace}\big)\big( {\textstyle \frac12}\,d^{bde} + {\textstyle \frac{i}{3}}\,f^{bde}\big)
\nonumber\\
&+& g_4\,(p_-^ip_-^j)_{(2)}\cdot(\sigma_1^i\sigma_2^j)_{(2)}\,\big( {\textstyle \frac12}\,d^{ace} + {\textstyle \frac{i}{3}}\,f^{ace}\big)\big( {\textstyle \frac12}\,d^{bde} + {\textstyle \frac{i}{3}}\,f^{bde}\big)\,,
\end{eqnarray}
and at the NNLO of the $1/N_c$ expansion takes form,
\begin{eqnarray}\label{NNLO-pot}
V_{\rm NNLO} &=& 9\,h_1\,p_+^2\delta_{\bar\chi_1\chi_1}\delta_{\bar\chi_2\chi_2}\,\delta^{cd}\delta^{bd}
+ \frac14\,h_2\,\vec\sigma_1\cdot\vec\sigma_2\,\delta^{cd}\delta^{bd}
+ \frac14\,h_3\,\vec\sigma_1\cdot\vec\sigma_2\,i^2\,f^{ace}\,f^{bde}
+ h_4\,p_+^2\,i^2\,f^{ace}\,f^{bde}\,\delta_{\bar\chi_1\chi_1}\delta_{\bar\chi_2\chi_2}
\nonumber\\
&+& h_5\,p_+^2\,\vec\sigma_1\cdot\vec\sigma_2\,\big( {\textstyle \frac12}\,d^{ace} + {\textstyle \frac{i}{3}}\,f^{ace}\big)\big( {\textstyle \frac12}\,d^{bde} + {\textstyle \frac{i}{3}}\,f^{bde}\big)
+ \frac32\,i\,h_6\,(\vec p_+\times\vec p_-)\cdot(\vec\sigma_1 + \vec\sigma_2)\,\delta^{cd}\delta^{bd}
\nonumber\\
&+& i\,h_7\, (\vec p_+\times\vec p_-)\cdot \Big[ \vec\sigma_1\,\big( {\textstyle \frac12}\,d^{ace} + {\textstyle \frac{i}{3}}\,f^{ace}\big)\,i\,f^{bde}
+ \vec\sigma_2\,i\,f^{ace}\,\big( {\textstyle \frac12}\,d^{bde} + {\textstyle \frac{i}{3}}\,f^{bde}\big) \Big]
\nonumber\\
&+& \frac32\,i\,h_8\,(\vec p_+\times\vec p_-)\cdot(\vec\sigma_1 + \vec\sigma_2)\,i^2\,f^{ace}\,f^{bde}
+ \frac14\,h_9\,(p_-^ip_-^j)_{(2)}\cdot(\sigma_1^i\sigma_2^j)_{(2)}\,\delta^{cd}\delta^{bd}
\nonumber\\
&+& \frac14\,h_{10}\,(p_-^ip_-^j)_{(2)}\cdot(\sigma_1^i\sigma_2^j)_{(2)}\,i^2\,f^{ace}\,f^{bde}
+ h_{11}\,(p_+^ip_+^j)_{(2)}\cdot(\sigma_1^i\sigma_2^j)_{(2)}\,\big( {\textstyle \frac12}\,d^{ace} + {\textstyle \frac{i}{3}}\,f^{ace}\big)\big( {\textstyle \frac12}\,d^{bde} + {\textstyle \frac{i}{3}}\,f^{bde}\big).
\end{eqnarray}
We note that the $N_c$ scales of the above potentials are $V_{\rm LO} \sim N_c$ and $V_{\rm NNLO} \sim N_c^{-1}$\,.

By using Eqs. (\ref{pot-1}), (\ref{pot-2}), (\ref{pot-3}), (\ref{LO-pot}) and (\ref{NNLO-pot}), the $N_c$ scaling relations of the LECs can be extracted,
\begin{eqnarray}\label{LECs-Nc}
C_{1,2}^{(1)}\,\sim C_{1,2}^{(2)}\,\sim C_{1,2}^{(3)}\,\sim N_c\,,\quad\qquad C_{3,4,5}^{(1)}\,\sim C_{3,4,5}^{(2)}\,\sim C_{3,4,5}^{(3)}\,\sim N_c^{-1}\,,
\end{eqnarray}
where $\Lambda \sim N_c^0$ \cite{Kaplan:1996rk,Phillips:2014kna,Schindler:2015nga} is impled. Note that the couplings $C_{1,2,3}^{(1)}$\,, $C_{1,2,3,4,5}^{(2)}$\,, $C_{1,2,3}^{(3)}$ are LO of order $N_c$\, while the $N_c$ scaling of the $C_{3,4,5}^{(1)}$\,, $C_{3,4,5}^{(2)}$\, and $C_{3,4,5}^{(3)}$ are further suppressed by order $1/N_c^2$\,. We found that there is no NLO of the LECs in the $1/N_c$ expansion.% which is agreement with the $1/N_c$ expansion of NN potential in SU(2) flavor symmetry case.

Matching the spin and flavor structures between the octet-octet baryon potential of the SU(3) chiral Lagrangian and the $1/N_c$ expansion up to NNLO, the large-$N_c$ operator analysis leads to the relations between the LECs of the SU(3) baryon contact interaction and we find the following results,
\allowdisplaybreaks
\begin{eqnarray}
C_1^{(2)} &=& C_1^{(1)} + g_2 - 4\, h_4 \,\Lambda^2 \,,
\nonumber\\
C_2^{(2)} &=& C_2^{(1)} + g_2 + 4\, h_4\,\Lambda^2\,,
\nonumber\\
C_3^{(2)} &=& C_3^{(1)}-\frac{1}{2}\,g_2 +\frac{1}{8}\,h_3 - 4\,h_4 \,\Lambda^2 + 2\,h_+\,\Lambda^2\,,
\nonumber\\
C_4^{(2)} &=& C_4^{(1)}-\frac{1}{2}\,g_2 - \frac{3}{8}\,h_3 - 4\,h_4\,\Lambda^2 + 2\,h_+\,\Lambda^2 \,,
\nonumber\\
C_5^{(2)} &=& C_5^{(1)} + \frac{1}{4}\,h_3 + 4\,h_4\,\Lambda^2 - 4\,h_+\,\Lambda^2 + 2\,h_{10}\,\Lambda^2\,,
\nonumber\\
C_1^{(3)}&=& -\frac{1}{3}\,C_1^{(1)} +\frac{9}{2}\,g_1 - \frac{1}{3}\,g_2 - 18\,h_1\,\Lambda^2 + \frac{4}{3}\,h_4\,\Lambda^2\,,
\nonumber\\
C_2^{(3)}&=& -\frac{1}{3}\,C_2^{(1)} +\frac{9}{2}\,g_1 -\frac{1}{3}\,g_2 + 18\,h_1\,\Lambda^2 -\frac{4}{3}\,h_4\,\Lambda^2\,,
\nonumber\\
C_3^{(3)}&=&-\frac{1}{3}\,C_3^{(1)} -\frac{9}{4}\,g_1 +\frac{1}{6}\,g_2 - 18\,h_1\,\Lambda^2 + \frac{1}{16}\,h_2 - \frac{1}{24}\,h_3
+\frac{4}{3}\,h_4\,\Lambda^2 + \frac{3}{2}\,h_6\,\Lambda^2 -\frac{2}{3}\,h_+\,\Lambda^2\,,
\nonumber\\
C_4^{(3)}&=&-\frac{1}{3}\,C_4^{(1)} -\frac{9}{4}\,g_1 +\frac{1}{6}\,g_2 - 18\,h_1\,\Lambda^2 - \frac{3}{16}\,h_2 +\frac{1}{8}\,h_3
+\frac{4}{3}\,h_4\,\Lambda^2 + \frac{3}{2}\,h_6\,\Lambda^2-\frac{2}{3}\,h_+\,\Lambda^2\,,
\nonumber\\
C_5^{(3)}&=&-\frac{1}{3}\,C_5^{(1)} + 18\,h_1\,\Lambda^2 +\frac{1}{8}\,h_2 -\frac{1}{12}\,h_3 -\frac{4}{3}\,h_4\,\Lambda^2 -3\,h_6\,\Lambda^2 + h_9\,\Lambda^2 + \frac{4}{3}\,h_+\,\Lambda^2 - \frac{2}{3}\,h_{10}\,\Lambda^2\,,
\end{eqnarray}
where $h_+ = 2\,h_7/3 + 3\,h_8$\,. Note that the Jacobi identities for the $f$ and $d$ symbols,
\begin{eqnarray}
&& f^{abe}\,f^{ecd} + f^{bce}\,f^{ead} + f^{cae}\,f^{ebd} = 0\,,
\nonumber\\
&&\, d^{abe}\,f^{ecd} + d^{bce}\,f^{ead} + d^{cae}\,f^{ebd} = 0\,
\end{eqnarray}
have been used in the matching procedure.

To the LO contributions of the $1/N_c$ expansion, one can reduce the number of the free parameters with $\mathcal{O}\big( 1/N_c^2 \big)$ $\equiv$ $h_i$ corrections.
% by dropping the  $h_i$ couplings.
9 sum rules of the LECs of the SU(3) octet-octet baryon contact interactions in the ChEFT are derived
\begin{eqnarray}\label{LO-sumrules}
&& C_1^{(1)}= C_1^{(2)} = -3\,C_1^{(3)} - 2\,C_4^{(2)} - 6\,C_4^{(3)}
\,,\qquad C_2^{(1)}= C_2^{(2)} = -3\,C_2^{(3)} - 2\,C_4^{(2)} - 6\,C_4^{(3)}
\,,
\nonumber\\
&&C_3^{(1)}= C_3^{(2)} = -3\,C_3^{(3)} + \,C_4^{(2)} + 3\,C_4^{(3)}\,,\qquad C_4^{(1)}= C_4^{(2)}\,,\qquad C_5^{(1)} = C_5^{(2)}=-3\,C_5^{(3)}\,.
\end{eqnarray}%
%One finds 8 sum rules at the NLO of the $1/N_c$ expansion, the sum rules are given by
%\begin{eqnarray}\label{NLO-sumrules}
%&& C_1^{(2)}= -3\,C_1^{(3)}\,,\qquad C_2^{(2)}=C_2^{(1)}\,,\qquad C_3^{(2)}=C_3^{(1)}\,,\qquad C_4^{(2)}=C_4^{(1)}\,,\qquad C_5^{(2)}= -3\,C_5^{(3)}\,,
%\nonumber\\
%&& C_2^{(3)}=-\frac{1}{3}\,C_2^{(1)}\,,\qquad C_3^{(3)}=-\frac{1}{3}\,C_3^{(1)}\,,\qquad
%C_4^{(3)}=-\frac{1}{3}\,C_4^{(1)}\,.
%\end{eqnarray}
We find that there are 6 free parameters of the SU(3) octet-octet baryon contact interactions in the ChEFT from the large-$N_c$ operator analysis. At $N_c=3$, these sum rules are held up to corrections of the $1/N_c^2\approx 10\%$ approximately. In order to see the application of the 9 large-$N_c$ sum rules, we will apply our results to YN interactions in next section.

\section{Application of the large-$N_c$ sum rules to the J\"{u}lich hyperon-nucleon contact interactions at the LO}
In this section, we will apply the large-$N_c$ sum rules to the J\"{u}lich hyperon-nucleon contact interactions at LO \cite{Polinder:2006zh}. The LO contact terms of the chiral Lagrangians in Eq. (\ref{chi-L}) with the large component of the baryon spinors have 6 free parameters. They read, \cite{Polinder:2006zh},
\begin{eqnarray}
&& C_S^{(1)}\,,\quad C_S^{(2)}\,,\quad C_S^{(3)}\,,\quad C_T^{(1)}\,,\quad C_T^{(2)}\,,\quad C_T^{(3)}\,.
\end{eqnarray}
%\begin{eqnarray}
%\mathcal{L}_{LO} &=& -\frac12\left[ C_S^{(1)}\left( \varphi_B^\dagger \varphi_B\right)\left( \varphi_B^\dagger \varphi_B\right)
%+ C_T^{(1)}\left( \varphi_B^\dagger\,\vec\sigma\, \varphi_B\right)\cdot\left( \varphi_B^\dagger\,\vec\sigma\, \varphi_B\right) \right]\mathcal{F}^{(1)}
%\nonumber\\
%&& -\frac12\left[ C_S^{(2)}\left( \varphi_B^\dagger \varphi_B\right)\left( \varphi_B^\dagger \varphi_B\right)
%+ C_T^{(2)}\left( \varphi_B^\dagger\,\vec\sigma\, \varphi_B\right)\cdot\left( \varphi_B^\dagger\,\vec\sigma\, \varphi_B\right) \right]\mathcal{F}^{(2)}
%\nonumber\\
%&& -\frac12\left[ C_S^{(3)}\left( \varphi_B^\dagger \varphi_B\right)\left( \varphi_B^\dagger \varphi_B\right)
%+ C_T^{(3)}\left( \varphi_B^\dagger\,\vec\sigma\, \varphi_B\right)\cdot\left( \varphi_B^\dagger\,\vec\sigma\, \varphi_B\right) \right]\mathcal{F}^{(3)} ,
%\end{eqnarray}
%where $\varphi_B$ is the large component of the baryon spinors and
%\begin{eqnarray}
%\mathcal{F}^{(1)} &=&  \frac13\,\delta^{cd}\delta^{ba} + \frac12\,\big(\,d^{cde} +if^{cde} \big)\big( d^{eba} + if^{eba}\big) ,
%\nonumber\\
%\mathcal{F}^{(2)} &=&  \frac13\,\delta^{ca}\delta^{db} + \frac12\,\big(\,d^{cae} +if^{cae} \big)\big( d^{edb} + if^{edb}\big) ,
%\nonumber\\
%\mathcal{F}^{(3)} &=&  \delta^{ca}\delta^{db}\,.
%\end{eqnarray}
The $C_{S,T}^{(1,2,3)}$ are linear combinations of the coupling constants in Eq. (\ref{chi-L}) as
\begin{eqnarray}
C_S^{(1,2,3)} = C_1^{(1,2,3)} + C_2^{(1,2,3)}\,,\qquad\quad  C_T^{(1,2,3)} = C_3^{(1,2,3)} - C_4^{(1,2,3)}\,.
\end{eqnarray}
The operator from the couplings, $C_5^{(1,2,3)}$ does not contribute to the YN potentials at the LO of the chiral expansion.
Applying the large-$N_c$ sum rules in Eq. (\ref{LO-sumrules}), we find 3 sum rules i.e.,
\begin{eqnarray}\label{LO-PW-sumrules}
C_S^{(1)} = C_S^{(2)}\,,\qquad  C_T^{(1)} = C_T^{(2)} = -3\,C_T^{(3)}\,.
\end{eqnarray}
Above sum rules give only 3 free parameters and the $N_c$ scalings of those parameters are given by
\begin{eqnarray}\label{C_ST-Nc}
C_S^{(1,2,3)}\sim N_c\,,\qquad C_T^{(1,2,3)}\sim N_c^{-1}\,.
\end{eqnarray}
%
%Up to the NLO of the $1/N_c$ expansion, the large-$N_c$ sum rules in Eq. (\ref{NLO-sumrules}) provide 3 sum rules
%\begin{eqnarray}\label{NLO-PW-sumrules}
%C_S^{(3)} = -\frac13\,C_S^{(2)}\,,\qquad C_T^{(2)} = C_T^{(1)}\,,\qquad C_T^{(2)} = -\frac13\,C_T^{(1)}\,.
%\end{eqnarray}
It is interesting to note that $N_c$ scalings of the $C_{S,T}^{(1,2,3)}$ in Eq. (\ref{C_ST-Nc}) agree with the NN case \cite{Kaplan:1996rk,Phillips:2013rsa}. The sum rules in Eq. (\ref{LO-PW-sumrules}) are useful for calculating the partial wave potentials at the LO in the chiral expansion of the hyperon-nucleon scattering. The hyperon-nucleon partial wave potentials at LO have been constructed and studied in Ref. \cite{Polinder:2006zh} and also re-investigated in \cite{Li:2016paq}.
According to the SU(3) flavor symmetry, the authors of the Ref. \cite{Polinder:2006zh} find that there are only 5 parameters (potentials) which are used to fit the experimental data of the hyperon-nucleon scattering. The parameters are read
\begin{eqnarray}\label{PW-free}
C_{1S0}^{\Lambda\Lambda} \equiv V_{1S0}^{\Lambda\Lambda}\,,\qquad C_{3S1}^{\Lambda\Lambda} \equiv V_{3S1}^{\Lambda\Lambda}\,,\qquad
C_{1S0}^{\Sigma\Sigma} \equiv V_{1S0}^{\Sigma\Sigma}\,,\qquad C_{3S1}^{\Sigma\Sigma} \equiv V_{3S1}^{\Sigma\Sigma}\,,\qquad
C_{3S1}^{\Lambda\Sigma} \equiv V_{3S1}^{\Lambda\Sigma}\,,
\end{eqnarray}
where the J\"{u}lich model of the LO hyperon-nucleon potentials are written in terms of the couplings $C_{S,T}^{(1,2,3)}$ in the following forms \cite{Polinder:2006zh},
\begin{eqnarray}\label{YN-LO}
V^{\Lambda\Lambda}_{1S0} &=& 4\pi\left[\frac{1}{6}\left(C^{(1)}_S-3C^{(1)}_T\right)+\frac{5}{3}\left(C^{(2)}_S-3C^{(2)}_T\right)+2\left(C^{(3)}_S-3C^{(3)}_T\right)\right],
\nonumber \\
V^{\Lambda\Lambda}_{3S1} &=& 4\pi\left[\frac{3}{2}\left(C^{(1)}_S+C^{(1)}_T\right)+\left(C^{(2)}_S+C^{(2)}_T\right)+2\left(C^{(3)}_S+C^{(3)}_T\right)\right],
\nonumber \\
V^{\Sigma\Sigma}_{1S0} &=& 4\pi\left[2\left(C^{(2)}_S-3C^{(2)}_T\right)+2\left(C^{(3)}_S-3C^{(3)}_T\right)\right]  ,
\nonumber \\
V^{\Sigma\Sigma}_{3S1} &=& 4\pi\left[-2\left(C^{(2)}_S+C^{(2)}_T\right)+2\left(C^{(3)}_S+C^{(3)}_T\right)\right]  ,
\nonumber \\
V^{\Lambda\Sigma}_{3S1} &=& 4\pi\left[-\frac{3}{2}\left(C^{(1)}_S+C^{(1)}_T\right)+\left(C^{(2)}_S+C^{(2)}_T\right)\right].
\label{eq:2.16}
\end{eqnarray}
Using the sum rules in Eq. (\ref{LO-PW-sumrules}) to the 5 free parameters in Eq. (\ref{PW-free}), one finds at LO of the $1/N_c$ expansion,
\begin{eqnarray}\label{s-wave-sumrules}
C_{1S0}^{\Sigma\Sigma} = \frac87\,C_{1S0}^{\Lambda\Lambda} - \frac17\,C_{3S1}^{\Lambda\Lambda} - \frac{11}{21}\,C_{3S1}^{\Lambda\Sigma} \,,\qquad
C_{3S1}^{\Sigma\Sigma} = C_{3S1}^{\Lambda\Lambda} + 9\,C_{3S1}^{\Lambda\Sigma}\,.
\end{eqnarray}
Note that all of the LECs has the same $N_c$ scaling as $N_c$.
%and up to NLO of the $1/N_c$ expansion,
%\begin{eqnarray}
%C_{1S0}^{\Lambda\Lambda} = \frac78\,C_{1S0}^{\Sigma\Sigma} +\frac{1}{48}\,C_{3S1}^{\Sigma\Sigma} -\frac19\,C_{3S1}^{\Lambda\Sigma} \,,\qquad
%C_{3S1}^{\Lambda\Lambda} = -\frac12\,C_{3S1}^{\Sigma\Sigma} - C_{3S1}^{\Lambda\Sigma}  \,.
%\end{eqnarray}
The large-$1/N_c$ analysis of the LO YN potentials predicts that there are 3 free parameters at the LO of $1/N_c$ expansion with $\mathcal{O}\big( 1/N_c^2\big)$ corrections. With the same manner of the large-$N_c$ analysis of the LO YN potentials, one can apply the sum-rules in Eq. (\ref{LO-sumrules}) for the partial-wave analysis in the YN potentials at NLO in Ref. \cite{Haidenbauer:2013oca} as well as for the YY sector in Refs. \cite{Polinder:2007mp,Haidenbauer:2015zqb,Haidenbauer:2013oca}.

%%%%%%%%%%%%%%%%%%%%%%%%%%%%%%%%%%%%%%%%%%%%%%%%%%%%%%%%%%%%%%%%%%%%%%%%%%%%%%%%%%%%%%%%%%%%%%%%%%%%%%%%%%%%%%%%%%%%%%%%%%%%%%%%%%%%%%%%%
\begin{table}
%\centering
    \begin{tabular}{l|cccccc}
  \hline\hline
  & $C^{\Lambda \Lambda}_{1S0}$ & $C^{\Sigma \Sigma}_{1S0}$ & $C^{\Lambda \Lambda}_{3S1}$ & $C^{\Sigma \Sigma}_{3S1}$ & $C^{\Lambda \Sigma}_{3S1}$ \\
  \hline
~~EG~~ & $-0.04795(151)$ & $-0.07546(81)$ & $-0.01727(124)$ & $0.36367(30310)$ & $0.01271(471)$\\
~~HB~~ & $-0.03894(1)$ & $-0.07657(1)$ & $-0.01629(13)$ & $0.20029(14050)$ &$-0.00176(304)$ \\
  \hline\hline
\end{tabular}
\caption{Best-fitted values of $YN$ s-wave LECs (in units of $10^4$ GeV$^{-2}$) for cut-off, $\Lambda=600$ MeV in the EG and HB approaches \cite{Li:2016paq}.}
\label{tab_LECs600}
\end{table}
%%%%%%%%%%%%%%%%%%%%%%%%%%%%%%%%%%%%%%%%%%%%%%%%%%%%%%%%%%%%%%%%%%%%%%%%%%%%%%%%%%%%%%%%%%%%%%%%%%%%%%%%%%%%%%%%%%%%%%%%%%%%%%%%%%%%%%%%%
Next we will compare the prediction of the large-$N_c$ sum rules in Eq. (\ref{s-wave-sumrules}) with the best fitted values of the LECs from YN scattering data in Ref. \cite{Li:2016paq}. This reference has performed the partial wave analysis of the YN s-wave scattering by using the same chiral Lagrangian as in our work. Authors in Ref. \cite{Li:2016paq} have used two approaches to solve scattering amplitudes via Kadyshevsky equation with the relativistic covariant ChEFT (referred as EG) and Lippmann-Schwinger equation with the heavy-baryon formalisms (referred as HB). The relativistic covariant ChEFT (EG) approach is also used to study NN interactions in \cite{Ren:2016jna}.  The best fitted values of the LECs are shown in Tab. \ref{tab_LECs600}. We will use the LECs, $C_{1S0}^{\Lambda\Lambda}$, $C_{3S1}^{\Lambda\Lambda}$ and $C_{3S1}^{\Lambda\Sigma}$ as input values in Eq. (\ref{s-wave-sumrules}) and the large-$N_c$ sum rules predict that
\begin{eqnarray}\label{LECs-sumrules}
&& C_{1S0,{\rm EG}}^{\Sigma\Sigma} = -0.06327\,,\qquad C_{3S1,{\rm EG}}^{\Sigma\Sigma} = 0.1271\,,
\nonumber\\
&& C_{1S0,{\rm HB}}^{\Sigma\Sigma} = -0.04333\,,\qquad C_{3S1,{\rm HB}}^{\Sigma\Sigma} = -0.0176\,.
\end{eqnarray}
Comparing the LECs, $C_{1S0}^{\Sigma\Sigma}$ and $C_{3S1}^{\Sigma\Sigma}$ from the large-$N_c$'s predictions with the best fitted values in Tab. \ref{tab_LECs600}, we found that $C_{1S0}^{\Sigma\Sigma}$ and $C_{3S1}^{\Sigma\Sigma}$ from large-$N_c$ are in the same order as the best fitted values and with the same relative sign in EG approach. On the other hand, for the HB formalisms, the $C_{1S0}^{\Sigma\Sigma}$ is also in the same order as the large-$N_c$ value and with the same relative sign. But for the $C_{3S1}^{\Sigma\Sigma}$ value in HB approach, it is different in order of magnitude of 1 with the large-$N_c$ prediction and with different relative sign. One notes that the LECs best fitted values from EG and HB approaches have statistical uncertainties at 68 \% (one sigma) level. While Ref. \cite{Li:2016paq} concluded that there is not much difference between two approaches. But the large-$N_c$ sum rules in this work can show that the LECs from EG approach is more consistent with the predictions of large-$N_c$ than the HB formalism.

\section{Conclusions}
In this work, we studied the large-$N_c$ operator analysis of the octet-octet baryon potential from the SU(3) ChEFT. The minimal set of the octet-octet baryon potential is derived by using the relativistic constraints as suggestion in Refs. \cite{Girlanda:2010ya,Girlanda:2010zz} as well as the Claley-Hamilton identity and Fierz rearrangement to eliminate the redundant operators as shown in Ref. \cite{Polinder:2006zh}. Up to NLO of $Q/\Lambda$ expansion, we found 27 operators for the octet-octet baryon potential in SU(3) flavor symmetry, 6 in LO and 21 in NLO of the small momentum scale.
%We found only operator at the NLO potential and it has order of $N_c^0$. In contrast to the $1/N_c$ expansion of the NN potential, there is no potential at order $N_c^0$ \cite{Kaplan:1996rk,Phillips:2013rsa}. This is because of the $N_c$ order of the flavor operator in SU(3) flavor symmetry has  $\big(B\,|\,T^a\,|\,B\big) \sim N_c$ while the isospin operator in SU(2) flavor symmetry scales like $\big(B\,|\,\tau^a\,|\,B\big) \sim N_c^0$.

The octet-octet baryon potential in the at LO in The $1/N_c$ expansion is of order $N_c$ and there are 4 operators while he NNLO potential is of order $1/N_c$ and we found 11 operators. The LECs of the ChEFT have two $N_c$ scalings, namely $N_c$ and $1/N_c$ orders as shown in Eq. (\ref{LECs-Nc}). Interestingly, the extension of the flavor symmetry from SU(2) to SU(3) in the large-$N_c$ operator analysis does not change the profile of the potential in terms of the $1/N_c$ expansion. There is no NLO for the SU(3) octet-octet baryon potential as for the NN potential \cite{Kaplan:1996rk,Phillips:2013rsa}.

The matching between the octet-octet baryon potential and the $1/N_c$ operator expansion leads to 6 free parameters of the LECs from the SU(3) chiral Lagrangian at the LO of the $1/N_c$ expansion with $\mathcal{O}\big( 1/N_c^2\big)\approx 10\%$ correction. The application of the sum rules in Eqs. (\ref{LO-sumrules}) from the lareg-$N_c$ constraint to the partial-wave potential of the YN interactions at LO of the chiral expansion reduces the LECs of the YN optential to 3 from 5.

The comparison of the large-$N_c$ predictions of the LECs with the best fitted values from the YN s-wave scattering reveals that the large-$N_c$ prediction of the LECs is more consistent with the EG results than the HB formalisms. Noted that The theoretical results from the EG and HB approaches in Ref. \cite{Li:2016paq} are quantitatively similar in describing the YN scattering experimental data.

The large-$N_c$ sum rules in this work can also be applied to the NLO of the YN interactions and extended to the ChEFT potential of the YY sector. In addition, we expect that future lattice QCD calculations may check the hierarchy of the $N_c$ scalings of the LECs and the large-$N_c$ sum rules predicted in this work.

\section*{Acknowledgments}
We would like to thank Daniel Phillips and Carlos Schat %Matthias Schindler
for carefully reading manuscript and useful comments. We also thank Li-Shen Geng for explaining detail of the best fitted values of LECs from YN scattering data. XL acknowledges support by National Natural Science Foundation of China (Project No. 11547182), and the Doctoral Scientific Research Foundation of Liaoning Province (Project No. 201501197). This work is partly supported by Thailand Research Fund (TRF) under contract No. MRG5980255 (DS). YY and DS acknowledge support from, Suranaree University of Technology (SUT) and the Office of the Higher Education Commission under NRU project of Thailand (SUT-COE: High Energy Physics \& Astrophysics). DS thanks Chamaipawn Jaipang for supporting useful references.

%%%%%%%%%%%%%%%%%%%%%%%%%%%%%%%%%%%%%%%%%%%%%%%%%%%%%%%%%%%%%%%%%%%%%%%%%%%%%%%%%%%%%%%%%%%%%%%%%%%%%%%%%%%%%%%%%%%%%%%%%%%%%%%%%%%%%%%%%%%%%%%%%%%%%%%%%%%
%%%%%%%%%%%%%%%%%%%%%%%%%%%%%%%%%%%%%%%%%%%%%%%%%%%%%%%%%%%%%%%%%%%%%%%%%%%%%%%%%%%%%%%%%%%%%%%%%%%%%%%%%%%%%%%%%%%%%%%%%%%%%%%%%%%%%%%%%%%%%%%%%%%%%%%%%%%%
\appendix
\section{The non-relativistic reductions of the chiral Lagrangian}\label{appA}
In this appendix, we derive the non-relativistic reductions of the chiral Lagrangian in Eq. (\ref{chi-L}). Here we follow the derivation from Ref. \cite{Girlanda:2010ya,Girlanda:2010zz} and focus for the spin (Dirac) structures of the chiral Lagrangian only. The chiral Lagrangian can be re-written in terms of operator as
\begin{eqnarray}\label{op-chi-L}
\widetilde{O}_1 &\equiv& (\bar{B} B) (\bar{B} B)\,,
\nonumber\\
\widetilde{O}_2 &\equiv& (\bar{B} \gamma_\mu B) ( \bar{B}\gamma^\mu B )\,,
\nonumber\\
\widetilde{O}_3 &\equiv& ( \bar{B} \sigma_{\mu\nu} B)( \bar{B} \sigma^{\mu\nu}B)\,,
\nonumber\\
\widetilde{O}_4 &\equiv& (\bar{B} \gamma_\mu \gamma_5 B)( \bar{B} \gamma^\mu \gamma_5 B )\,,
\nonumber\\
\widetilde{O}_5 &\equiv& (\bar{B} \gamma_5 B)(\bar{B} \gamma_5 B )\,.
\end{eqnarray}
\begin{table}
\begin{tabular}{c|c}
\hline
$O_S$  & $(\varphi_B^\dagger \varphi_B)(\varphi_B^\dagger \varphi_B)$ \\
$O_T$  & $(\varphi_B^\dagger {\bm {\sigma}} \varphi_B)\cdot
(\varphi_B^\dagger{\bm {\sigma}}\varphi_B)$ \\
\hline
$O_1$  &  $(\varphi_B^\dagger \overrightarrow{\bm \nabla} \varphi_B)^2
+{\rm h.c.}$   \\
$O_2$  & $(\varphi_B^\dagger \overrightarrow{\bm \nabla} \varphi_B )\cdot
( \varphi_B^\dagger \overleftarrow{\bm \nabla} \varphi_B) $\\
$O_3$  &  $(\varphi_B^\dagger \varphi_B) ( \varphi_B^\dagger \overrightarrow{\bm \nabla}^2 \varphi_B)+{\rm h.c.}$  \\
$O_4$  & $i \,( \varphi_B^\dagger \overrightarrow{\bm \nabla} \varphi_B) \cdot (\varphi_B^\dagger \overleftarrow{\bm \nabla}
  \times {\bm \sigma} \varphi_B )+ {\rm h.c.}$ \\
$O_5$ & $i \, (\varphi_B^\dagger \varphi_B)(\varphi_B^\dagger \overleftarrow{\bm \nabla}
\cdot {\bm \sigma} \times \overrightarrow{\bm \nabla} \varphi_B)$ \\
$O_6$ & $i \, (\varphi_B^\dagger {\bm \sigma} \varphi_B) \cdot
(\varphi_B^\dagger \overleftarrow{\bm \nabla} \times
\overrightarrow{\bm \nabla} \varphi_B)$\\
$O_7$  & $( \varphi_B^\dagger {\bm \sigma} \cdot
\overrightarrow{\bm \nabla} \varphi_B) (\varphi_B^\dagger {\bm \sigma}\cdot \overrightarrow{\bm \nabla} \varphi_B) +{\rm h.c.}$ \\
$O_8$  & $(\varphi_B^\dagger \sigma^j
\overrightarrow{\nabla^k} \varphi_B)(\varphi_B^\dagger \sigma^k \overrightarrow{\nabla^j} \varphi_B) + {\rm h.c.}$  \\
$O_9$  &   $(\varphi_B^\dagger \sigma^j
\overrightarrow{\nabla^k} \varphi_B)(\varphi_B^\dagger \sigma^j \overrightarrow{\nabla^k} \varphi_B) + {\rm h.c.}$ \\
$O_{10}$  & $(\varphi_B^\dagger {\bm \sigma} \cdot
\overrightarrow{\bm \nabla}\varphi_B)
(\varphi_B^\dagger \overleftarrow{\bm \nabla}\cdot {\bm \sigma} \varphi_B)$ \\
$O_{11}$  & $(\varphi_B^\dagger \sigma^j \overrightarrow{\nabla^k}\varphi_B)
(\varphi_B^\dagger \overleftarrow{\nabla^j} \sigma^k  \varphi_B)$ \\
$O_{12}$  &  $(\varphi_B^\dagger \sigma^j \overrightarrow{\nabla^k} \varphi_B)
(\varphi_B^\dagger \overleftarrow{\nabla^k} \sigma^j  \varphi_B)$\\
$O_{13}$  &  $ (\varphi_B^\dagger
\overleftarrow{\bm \nabla}\cdot{\bm  \sigma} \,\overrightarrow{\nabla^j} \varphi_B)
(\varphi_B^\dagger \sigma^j \varphi_B) +{\rm h.c.}$\\
$O_{14}$  & $2\, (\varphi_B^\dagger
\overleftarrow{\bm \nabla} \sigma^j \cdot \overrightarrow{\bm \nabla} \varphi_B) (\varphi_B^\dagger \sigma^j \varphi_B)$ \\
\hline
\end{tabular}
\caption{Operators of the LO and NLO contact
term interactions \cite{Ordonez:1996}, the left (right) arrow on $\nabla$ indicates that the gradient operates on the left (right) field.
Normal-ordering of the field operator products is implied.}\label{NR-op}
\end{table}
The relativistic fermion field $B(x)$ can be expanded to the positive energy components $\varphi_B(x)$ in the following from \cite{Girlanda:2010ya,Girlanda:2010zz},
\begin{eqnarray}
B(x) = \left[ \left(\begin{array}{c}
 1   \\
 0
\end{array} \right) - \frac{i}{2M} \left(\begin{array}{c}
 0   \\
 {\bm {\sigma}}\cdot {\bm {\nabla}}
\end{array} \right) + \frac{1}{8M^2}\left(\begin{array}{c}
 {\bm {\nabla}}^2   \\
 0
\end{array} \right)\right]\varphi_B(x) + \mathcal{O}\big( Q^3\big)\,,
\end{eqnarray}
where $M$ and $Q$ are baryon mass in SU(3) flavor symmetry limit and small momentum scale respectively. Up to order $Q^2$, the non-relativistic reductions of the operators in Eq. (\ref{op-chi-L}) are given by
\begin{eqnarray}\label{NR-reduce}
\widetilde{O}_1 &\stackrel{{\rm NR}}{\simeq}& O_S+\frac{1}{4 M^2} \left( O_1 + 2\, O_2 + 2\, O_3 + 2\, O_5 \right),
\nonumber\\
\widetilde{O}_2 &\stackrel{{\rm NR}}{\simeq}& O_S +\frac{1}{4 M^2}
  \left(-4\, O_2 -2\, O_5 +4\, O_6 +O_7 - O_9 + 2\, O_{10} - 2\, O_{12} \right),
\nonumber\\
\widetilde{O}_3 &\stackrel{{\rm NR}}{\simeq}& O_T+ \frac{1}{4 M^2} \left( -O_1 -
2\, O_2  - 4\, O_5 + 2\, O_6  + O_7 - 2\, O_8 +2\, O_{10} -4\, O_{12}
- 2\, O_{13} \right),
\nonumber\\
\widetilde{O}_4 &\stackrel{{\rm NR}}{\simeq}& -O_T -\frac{1}{4 M^2} \left(- 2 \, O_6 + O_7 - O_9 - 2 \, O_{10} - 2\, O_{12} + 2\,O_{13} - 2\, O_{14} \right),
\nonumber\\
\widetilde{O}_5 &\stackrel{{\rm NR}}{\simeq}& \frac{1}{4 M^2}\left(O_7
+2\, O_{10}\right),
\end{eqnarray}
where we took the above results from Refs. \cite{Girlanda:2010ya,Girlanda:2010zz} and the operators $O_i$ ($i=1,...,14$) are listed in Tab. \ref{NR-op}.

By using partial integrations, Ref.~\cite{Pastore:2009is} has been shown that there are only 12 operators are independent with the following constraints,
\begin{eqnarray}
O_7 + 2\,  O_{10} = O_8 + 2\, O_{11} \quad {\rm and} \quad O_4 + O_5 = O_6 \,.
\end{eqnarray}

Next step, one re-writes the non-relativistic reductions in Eq. (\ref{NR-reduce}) in terms of the basis in Eqs. (\ref{pot-1},\ref{pot-2},\ref{pot-3}) as \cite{Girlanda:2010ya},
\begin{eqnarray}
A_S &\equiv& \tilde O_S = O_S + \frac{1}{4M^2}\left( O_1 + O_3 + O_5 + O_6 \right),
\nonumber\\
A_T &\equiv& \tilde O_T = O_T - \frac{1}{4M^2}\left( O_5 + O_6 - O_7 + O_8 + 2\,O_{12} + O_{14} \right),
\nonumber\\
A_1 &\equiv& p_-^2\,\delta_{\bar\chi_1\chi_1}\delta_{\bar\chi_2\chi_2} = O_1 + 2\,O_2 \,,
\nonumber\\
A_2 &\equiv& p_+^2\,\delta_{\bar\chi_1\chi_1}\delta_{\bar\chi_2\chi_2} = 2\,O_2 + O_3 \,,
\nonumber\\
A_3 &\equiv& p_-^2\,\vec\sigma_1 \cdot \vec\sigma_2 = O_9 + 2\,O_{12} \,,
\nonumber\\
A_4 &\equiv& p_+^2\,\vec\sigma_1 \cdot \vec\sigma_2 = O_9 + O_{14} \,,
\nonumber\\
A_5 &\equiv& i \left(\vec p_+\times\vec p_- \right) \cdot (\vec\sigma_1 + \vec\sigma_2)/2 =  O_5 - O_6 \,,
\nonumber\\
A_6 &\equiv& (\vec p_-\cdot\vec\sigma_1)(\vec p_-\cdot\vec\sigma_2) = O_7 + 2\,O_{10}\,,
\nonumber\\
A_7 &\equiv& (\vec p_+\cdot\vec\sigma_1)(\vec p_+\cdot\vec\sigma_2) = O_7 + O_8 + 2\,O_{13}\,.
\end{eqnarray}
By using above relations, we obtain the non-relativistic reductions of the chiral Lagrangian in Eq. (\ref{chi-L}) in terms of the operators $A_i$ as,
\begin{eqnarray}
\widetilde{O}_1 &\simeq& A_S + \frac{1}{4 M^2}\left(A_2 - A_5\right),
\nonumber\\
\widetilde{O}_2 &\simeq& A_S -\frac{1}{4M^2}\left(A_1 + A_2 + A_3 - 3\,A_5 - A_6 \right),
\nonumber\\
\widetilde{O}_3 &\simeq& A_T -\frac{1}{4M^2}\left(A_1 + A_2 + A_3 - A_4 - 3\,A_5 - A_6 + A_7 \right),
\nonumber\\
\widetilde{O}_4 &\simeq& -A_T +\frac{1}{4M^2}\left(A_4 + A_5 + A_6 - A_7 \right),
\nonumber\\
\widetilde{O}_5 &\simeq& \frac{1}{4M^2}\,A_6 \,.
\end{eqnarray}

\end{document}